# Piezoelectric & Optical Set-up to measure an Electrical Field. Application to the Longitudinal Near-Field generated by a Tapered Coax

S. Euphrasie*, P. Vairac, B. Cretin and G. Lengaigne

Institut FEMTO-ST, Université de Franche-Comté, CNRS, ENSMM, 32 av. de l'Observatoire, F-25044 Besançon Cedex, France

* Corresponding author: sebastien.euphrasie@femto-st.fr

**Abstract:**

We propose a new set-up to measure an electrical field in one direction. This set-up is made of a piezoelectric sintered PZT film and an optical interferometric probe. We used this set-up to investigate how the shape of the extremity of a coaxial cable influences the longitudinal electrical near-field generated by it. For this application, we designed our set-up to have a spatial resolution of 100 μm in the direction of the electrical field. Simulations and experiments are presented.

# I. Introduction:

The most common method to measure an electrical field is with an antenna. If one wants to have a good spatial resolution, short electrical dipoles can be used. One of the drawbacks is the presence of metal (of the antenna but also of the connecting wires) that will disturb the electric field to be measured. When the spatial resolution is essential, the possibility that the wires can pick up the field is also a major drawback. To limit this effect, a change of frequency can be made (heterodyning). For instance, to measure the Specific Absorption Rate (SAR), short dipoles are usually connected to diodes that will convert the radio-frequency (RF) signal into continuous (DC) signal. High impedance wires, almost invisible to the RF signal, then transmit the information [1-2]. However, all frequency and phase information are lost. Another solution is to use optical systems, the information is then carried by an electromagnetic signal at a much higher frequency than the electrical field's one. We can find in the literature set-ups with an electro-optic crystal to change the polarisation of a laser beam according to the measured field [3]. Spatial resolution is approximately the size of the laser spot and the crystal's, which typically is 1mm x 1mm x 1mm. This sensor is sensitive only to one direction of the electrical field, depending of the orientation of the crystal.

We propose here another set-up using dielectrics and optics with no metal to perturb the measured field. It consists of a piezoelectric sintered PZT film and an optical heterodyne probe. The information is firstly converted into mechanical deformation, then into phase modulation of a laser beam. The measure depends only on the electrical field in one direction with a good spatial resolution in this direction. To demonstrate the feasibility of this sensor, we chose to investigate, as an application, the influence of using a conical extremity on the longitudinal electrical near-field generated by a coaxial cable. Considering that the highest frequency used is 500 MHz, the wavelength in air is greater than 0.6 m, much higher than the distance from the source. Therefore it is a near-field problem where non-travelling waves must be taken into account.

Coaxial cables, with or without the central conductor protruding, are used as antennas to measure electrical near-field [4-6]. Depending on the length of the protruding connector or the shape of the cable's apex – conical extremity or not – the sensitivity to the longitudinal electrical field can greatly vary. As far as we know, no study has been carried out concerning this topic, which is the reason we chose this application. Since emitting and receiving are similar in an antenna, our study concentrates on the emitting issue. We used "large" coaxial cables since it was easier to work with; but, as the problem can easily be scaled down, this study can be applied to MEMS, in particular to the structures developed in reference [7]. In order not to mix two different issues, we did not use the tip effect by sharpening the inner conductor as in reference [4].

We will first present this piezo-optical sensor, then the simulations concerning the conical coaxial cables and we will compare them with experimental data.

## II. Electrical Near Field Sensor

In order to measure the longitudinal electrical field $E_z$, we chose to convert it into displacement with a piezoelectric element and to measure the movement with a home-made heterodyne probe [8-10]. In optimal conditions, the sensitivity of the probe is 2 fm $/\sqrt{Hz}$. One of the advantages is that it is a non-contact method. The piezoelectric element must not be a mono-crystal so that the other components of the field will not induce a displacement. Using a sintered ceramic, the rotational symmetry along the polarization axis leads to piezoelectric coefficients $d_{13}$ and $d_{23}$ null. With the matrix notation, the piezoelectric equation is then:

$$\begin{bmatrix} S_1 \\ S_2 \\ S_3 \\ S_4 \\ S_5 \\ S_6 \end{bmatrix} = \begin{bmatrix} s_{11}^E & s_{12}^E & s_{13}^E & 0 & 0 & 0 \\ s_{12}^E & s_{11}^E & s_{13}^E & 0 & 0 & 0 \\ s_{13}^E & s_{13}^E & s_{33}^E & 0 & 0 & 0 \\ 0 & 0 & 0 & s_{44}^E & 0 & 0 \\ 0 & 0 & 0 & 0 & s_{44}^E & 0 \\ 0 & 0 & 0 & 0 & 0 & 2(s_{11}^E - s_{12}^E) \end{bmatrix} \begin{bmatrix} T_1 \\ T_2 \\ T_3 \\ T_4 \\ T_5 \\ T_6 \end{bmatrix} + \begin{bmatrix} 0 & 0 & d_{31} \\ 0 & 0 & d_{31} \\ 0 & 0 & d_{33} \\ 0 & d_{15} & 0 \\ d_{15} & 0 & 0 \\ 0 & 0 & 0 \end{bmatrix} \begin{bmatrix} E_1 \\ E_2 \\ E_3 \end{bmatrix} \quad (1)$$

where $S_i$, $s_{ik}^E$, $T_k$, $d_{ij}$ and $E_j$ respectively represent the strain, the compliance, the stress, the piezoelectric coefficients and the electrical field in the ceramic. If we neglect the clamping effects, since the piezoelectric ceramic can move freely there is no stress, negating the first term of the equation. We will also neglect the torsion due to $d_{15}$ and $d_{24}$ coefficients. Hence, with these approximations, the displacement along the polarization axis only comes from an electrical field along this axis. The strain is given by equation (2) :

$$S_3 = d_{33} \cdot E_3 \quad (2)$$

If there is no surface charge, the relation between the field in the ceramic $E_3$ and in the air $E_{3ext}$ is:

$$E_3 = E_{3ext} / \varepsilon_r \quad (3)$$

with $\varepsilon_r$ the relative dielectric constant of the ceramic. Therefore the displacement $\delta$ can be expressed as:

$$\delta = S_3 \cdot t = d_{33} \cdot E_{3ext} / \varepsilon_r \cdot t \quad (4)$$

where t is the thickness of the ceramic.

The spatial resolution of this sensor is given by its dimensions. Obviously, the field is "integrated" in its thickness, which will give an average of the field. Since the displacement is proportional to the thickness, a compromise must be made. For this study, we decided to work on a "large" coaxial antenna (the diameter of the outer conductor is around 60mm and the inner's one is 1mm), for simplicity's sake and because everything can be easily scaled down. Therefore, a thickness of 100 microns seems a good compromise: it is ten times smaller than the inner diameter, the displacement will be large enough and it is easy to fabricate. The lateral resolution depends on the size of the laser spot and the lateral dimension of the ceramic because of $d_{31}$ and clamping effect. We chose a square a little larger than the laser spot of the heterodyne probe without focalisation, that is to say 1mm x 1mm. The lateral resolution is then roughly the same size.

This sensor was fabricated with the following steps. A PZT plate was glued to a glass wafer with SU8 2002 photoresist. The PZT and the glass were then thinned to respectively 100 and 200 microns by mechanical polishing. Particular care was given to the PZT in order to have a mirror quality polishing. Cantilevers of 1 mm width were cut out with a diamond saw. At the same time, the PZT was cut in squares of 1 mm². The PZT was then mechanically removed except at the end of the cantilever. The cantilever was then glued to a set-up with two micrometric translations and two micrometric rotations so that the PZT reflects the probe's laser beam in optimal conditions.

Considering a piezoelectric coefficient $d_{33}$ of 290 pm/V, a relative dielectric constant $\varepsilon_r$ of 735, the displacement should be $4.10^{-17}$ m for $E_{3ext} = 1V/m$. In our case, the probe has a sensitivity of $5.10^5$ V/m, which gives $2.10^{-11}$V for $E_{3ext} = 1V/m$. Considering that in these conditions, with a lock-in amplifier, the noise level is about 20 nV, the threshold of detection of the electrical field in the air is around 1KV/m. This low value can be greatly enhanced in several ways. First, we can improve the sensitivity of the interferometric probe. The easiest possibilities are to use a more powerful laser and to change the optics to focus the beam. Using a cavity can also greatly increase the sensitivity. The electro-mechanical part can also be improved. Choosing a material with a higher $d_{33}/\varepsilon_r$ ratio can also lead to a slight increase in sensitivity. Mechanical resonances can also be used: The sensitivity of the sensor is then greatly increased in some frequency bands. Nonetheless, a calibration for each frequency of interest must be done for quantitative measurement. Since, in our application, our sensor serves only to compare two antennas, no calibration has been carried out. Finally, a dielectric mirror coating on the piezoelectric ceramic would increase the reflectance and therefore the power of the returning laser beam and the signal to noise ratio.

In order to have a better spatial resolution, the size of this sensor could be reduced, using photoresist mask and PZT sputtering for instance. Nonetheless, one must not forget that the sensitivity is proportional to the thickness. In order not to decrease the sensitivity by improving the lateral resolution, the laser beam must be focused.

## III. Simulations

The simulations of the coaxial cables were carried out with CST Microwave Studio, a commercial software using Finite Integration Technique (FIT) and the Perfect Boundary Approximation (PBA$^{TM}$) [11-12]. The FIT can be seen as a generalization of the FDTD (Finite Difference Time Domain) method. It discretizes the integral form of Maxwell's equations, rather than the differential one, on a pair of dual interlaced discretization grids. Electric voltages $\hat{e}$ and magnetic fluxes $\hat{b}$ are defined on the edges and facets of the primary grid. Magnetic voltages $\hat{h}$ and electric fluxes $\hat{d}$ are defined on the edges and faces of the secondary grid. For instance, with $C$ and $\tilde{C}$ the matrixes of the primary and dual grid respectively, Faraday's and Ampère's Grid Equations become:

$$C\hat{e} = -\frac{d}{dt}\hat{b} \quad \text{and} \quad \tilde{C}\hat{h} = \frac{d}{dt}\hat{d} + \hat{j} \qquad (5), (6)$$

with $\hat{j}$ the current density vector.

The PBA$^{TM}$ allows us to avoid the disadvantage of the staircase approximation of complex boundaries.

A schema of different antennas simulated is given in figure 1. The evolution of the longitudinal electrical field as a function of the distance z from the end of the coaxial cable was simulated for different shapes and frequencies – 50 MHz and 0.5 GHz are plotted in figures 2 & 3. The radius of the external conductor is 30 mm and the internal one is 0.5 mm. To insure that there is no point effect at the end of the outer conductor, a small ring of metal (0.5 mm thick) was added in the case of the shapes "cone2" and "expo1" (see figure 1). The power at the waveguide port is 1 W. The simulations were 3D with open boundaries and no symmetry. A parallelepiped of air (4 x 4 x 2 mm) was added in front of the tip of the antenna to refine the meshing. Two millimeters of background air were added in every direction (except at the port). Figure 2 clearly shows that having a tapered coaxial cable (cone1 & 2, expo1 & 2) increases the value of the longitudinal field "near" it (i.e. z < 0.8 mm in our case; this distance depends on the radius of the inner conductor). This effect is probably caused by two main reasons. Firstly, as the distance between the two metal decreases, the electrical field in the coaxial cable obviously increases at the tip. Secondly, the electrical field is perpendicular to the metal next to it, since the metal at the tip is tilted, this will increase the longitudinal field. This benefit decreases as the distance increases and, after some distance, the longitudinal field magnitude is lesser with a tapered extremity. Of course, in that case, the spatial resolution remains better (comparable to the size of the extremity), but that is not the topic of this article. At the frequency of 50 MHz, whichever tapered shape gives about the same longitudinal field. On the contrary, when the frequency is higher (for instance 0.5 GHz) and the wavelength is close to the dimensions of the cable, the shape is critical (cf. fig. 3). The shape "expo 1" gives the best result (i.e. higher field near the apex) followed by the cone with an angle about 45° (shape "cone2"). Removing the ring preventing the tip effect hardly changes the results for "expo1" and improves "cone2" so that the field is similar to that of

expo1. The conical shape with an angle about 45° will be used for the other simulations and experiments because of its simplicity.

The influence of the length of the protruding part of the inner conductor has also been simulated. In figure 4, the longitudinal field $E_z$ has been plotted against the distance z from the end of the central conductor, for different lengths p of the protuberance (p = 0 mm, 5 mm and 10 mm). Once again, the radius of the external conductor is 30 mm, the internal one is 0.5 mm, the length is 30 mm (therefore the angle is about 45°) and the frequency is 50 MHz. Figure 4 shows that having a protuberance slightly decreases the longitudinal field near the end of the cable, i.e. when the cone-shape increases it. The same study at higher frequencies (f = 0.5 GHz) shows the same results. We will therefore not use a protruding inner conductor for the following simulations and experiments.

The influence of the outer conductor's radius, with the inner conductor's radius fixed at 0.5 mm, was also investigated. Three radii were considered: 30mm, 15 mm and 5 mm. In each case, the angle was exactly 45°. The ratio between the cone and the coaxial cable of the same size were plotting versus the distance z in figure 5. The frequency of this plot is 5 MHz, but the same tendencies are obtained at 0.5 GHz. There is only a slight difference between the curves; but the larger the outer radius is, the higher the ratio is.

## IV. Experimental set-up and results

In the experimental set-up, we used two coaxial cables as emitting antennas. One had an abrupt ending, the other a conical extremity. The coaxial cables were quite "large" – the outer radius was 26 mm and the inner conductor had a radius of 0.85 mm – in order to have a better spatial resolution compared to the size of the coaxial cables. The dielectric was made of PVC, drilled in the centre to insert a copper wire and coated with silver paint to make the outer conductor. The two antennas were connected to a classical coaxial cable with a tapered coaxial transition made likewise and ending with a BNC connexion (cf. fig. 6).

They were mounted using three micrometric translation stages for x-y-z displacements and the three rotations could be adjusted and blocked manually with screws. They were the moving part of the experimental set-up for the field mapping. A low frequency generator was used as a power supply for the antennas and for the lock-in detection made by a DSP lock-in amplifier SR850 (Stanford Research Systems). The voltage supplied by the generator to the antennas was 10Vpp at 25 kHz. The antennas were centred compared to the piezoelectric element and were moved along the longitudinal axis z. The origin of the axis was taken in the centre of the piezoelectric element.

In figure 7, experimental results and simulations are plotted together. The simulations used a lossy dielectric (polyimide: $\varepsilon_r$ = 3.5, tangent delta = 0.003). Since there was no calibration, the simulations were re-scaled, both with the same factor. The experiment and simulation fit well together for the cone-shaped coaxial cable. Although that is not the case with the non-tapered coax, we observe the same tendency as in the simulations: the longitudinal field $E_z$ is greater with a cone-shaped coax when measured near the end of the cable (distance z less than about 1 mm); further it is smaller.

## V. Conclusions:

We proposed a new way to measure an electrical field in one direction with a good spatial resolution. The set-up is made of a piezoelectric sintered PZT film - in movement because of the electrical field - and an optical interferometric probe. The other components of the field have almost no effect. There is no perturbation limiting the spatial resolution because of the change of medium (mechanical then optical) which transmits the information. As a demonstration of the validity of this concept, we chose to use it to investigate the influence of a conical extremity on the longitudinal electrical near-field generated by a large coaxial cable. Simulations and experiments agreed: the nearer to the extremity of the cable, the higher is the longitudinal field of the tapered coaxial cable compared to a straight-shaped one. Several ways to improve the sensitivity of this set-up were suggested.

**Acknowledgement:**

The authors would like to acknowledge the help of David Vernier, Sylvain Ballandras, Ludovic Gauthier-Manuel, Jean-Claude Baudouy and Pierre Berthelot.

**Figures:**

**Figure 1:** Drawings of different shapes used in the simulations.

**Figure 2:** Simulation of the longitudinal electrical field $E_z$ versus the distance z from the end of the cable for different shapes at a frequency of 50 MHz.

**Figure 3:** Simulation of the longitudinal electrical field $E_z$ versus the distance z from the end of the cable for different shapes at a frequency of 0.5 GHz.

**Figure 4:** Simulation of the longitudinal electrical field $E_z$ versus the distance z from the end of the cable for different protruding length of the inner conductor at a frequency of 50 MHz.

**Figure 5:** Simulation of the ratio of longitudinal electrical field $E_z$ generated by a cone by a coax's one versus the distance z from the end of the cable for different outer conductor's radius at a frequency of 50 MHz. The radius of the inner conductor is 0.5 mm and the angles of the cones are 45°.

**Figure 6:** Experimental set-up

**Figure 7:** Comparison of experiment and simulation results for the longitudinal electrical field for the cone-shaped and the non-tapered coaxial cables versus the distance z from the end of the cable.

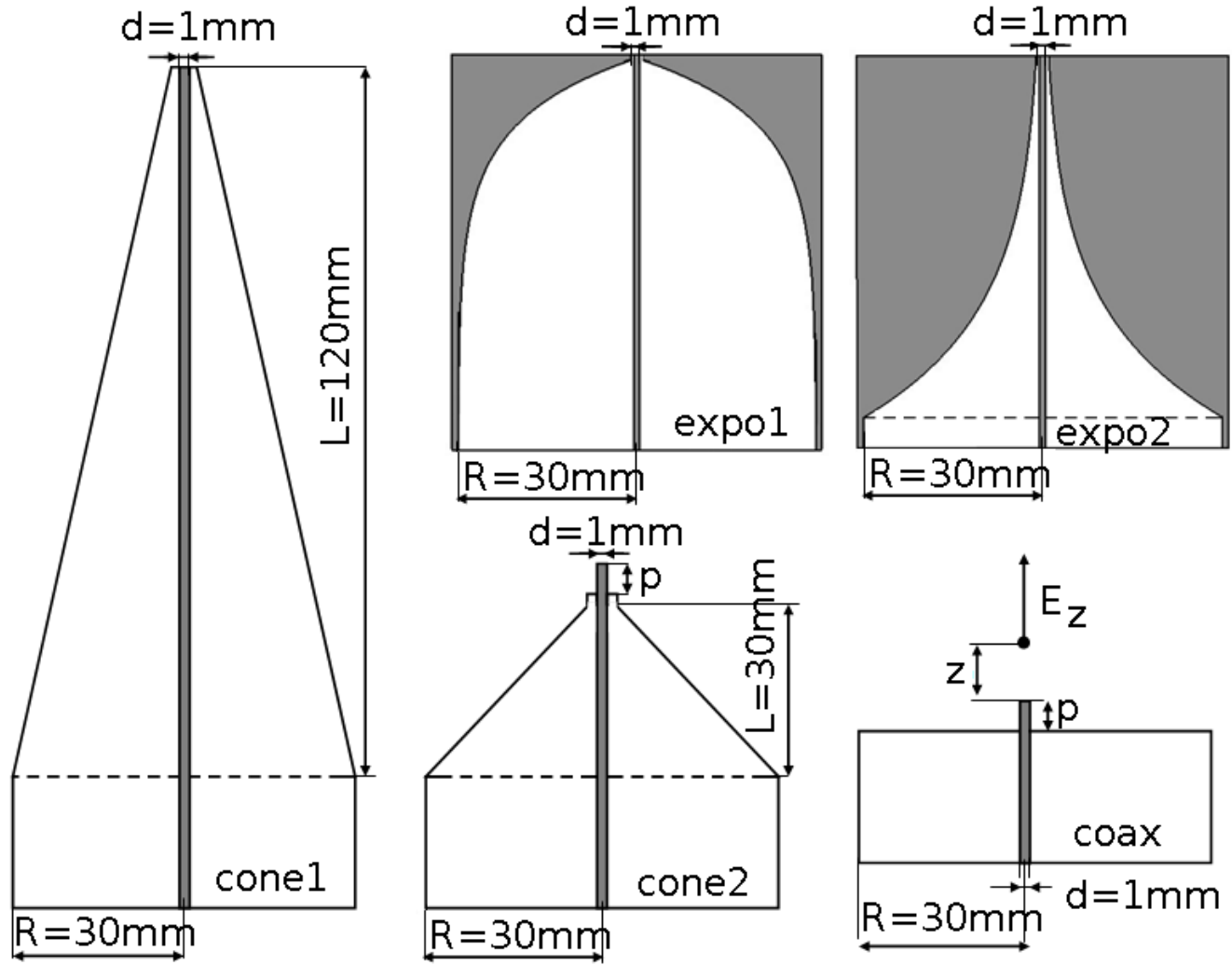

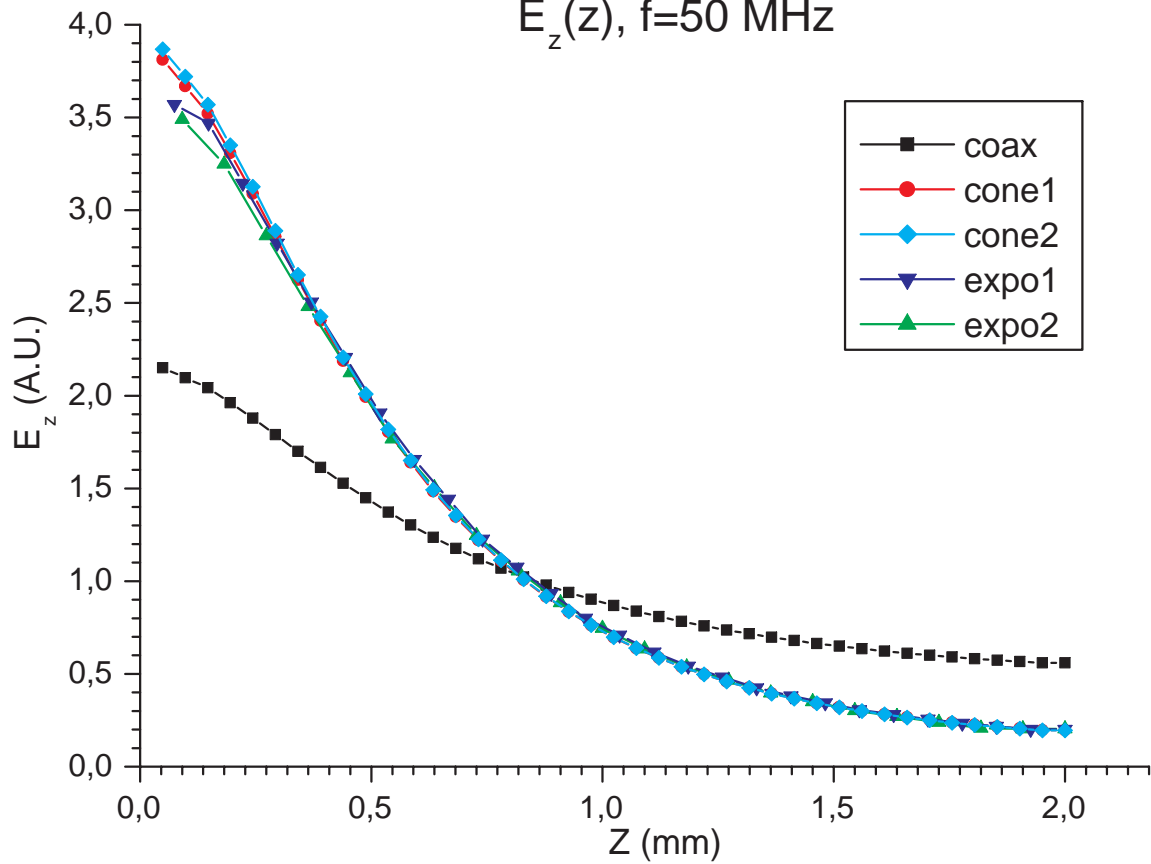

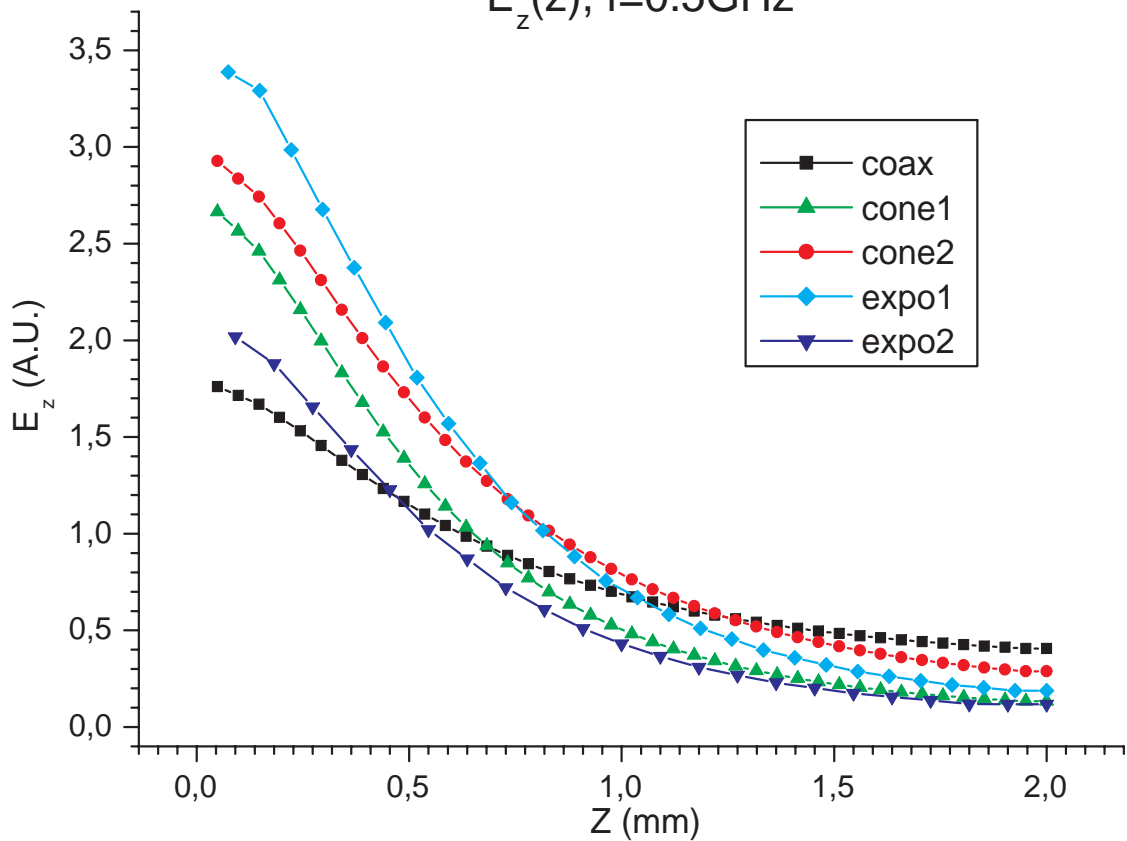

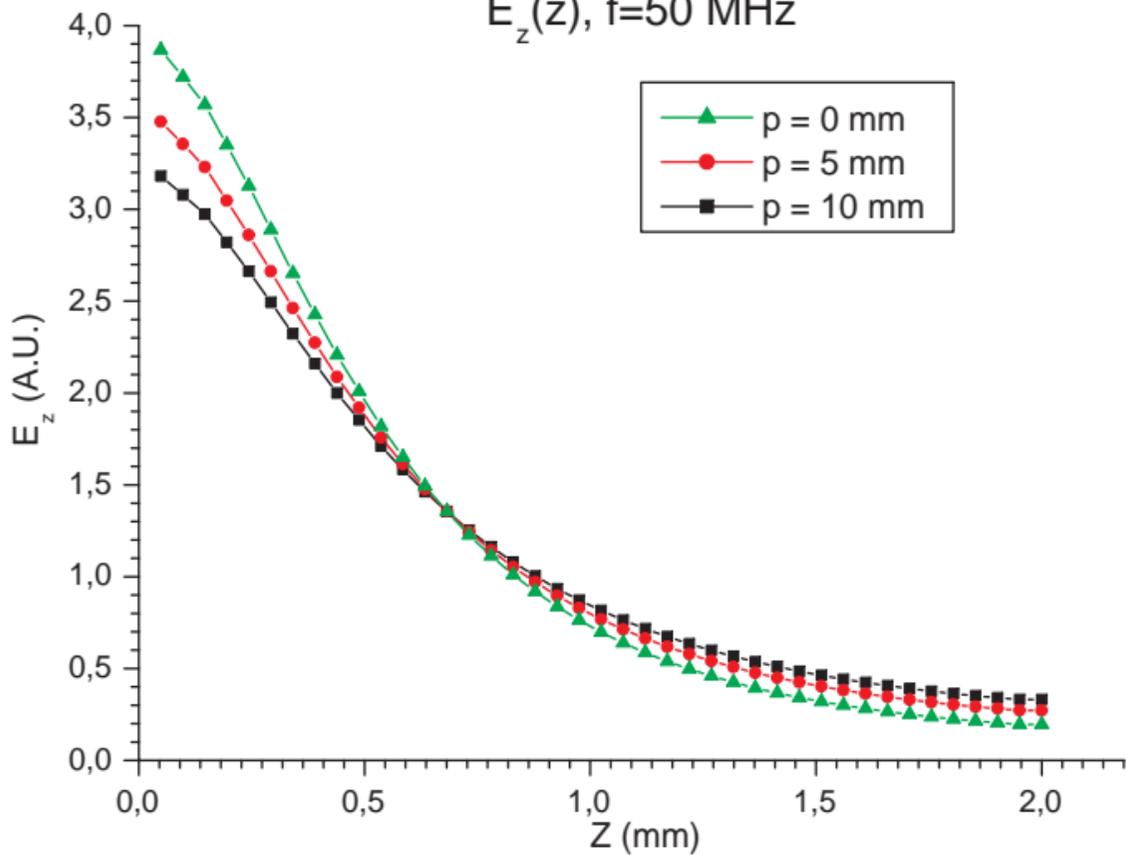

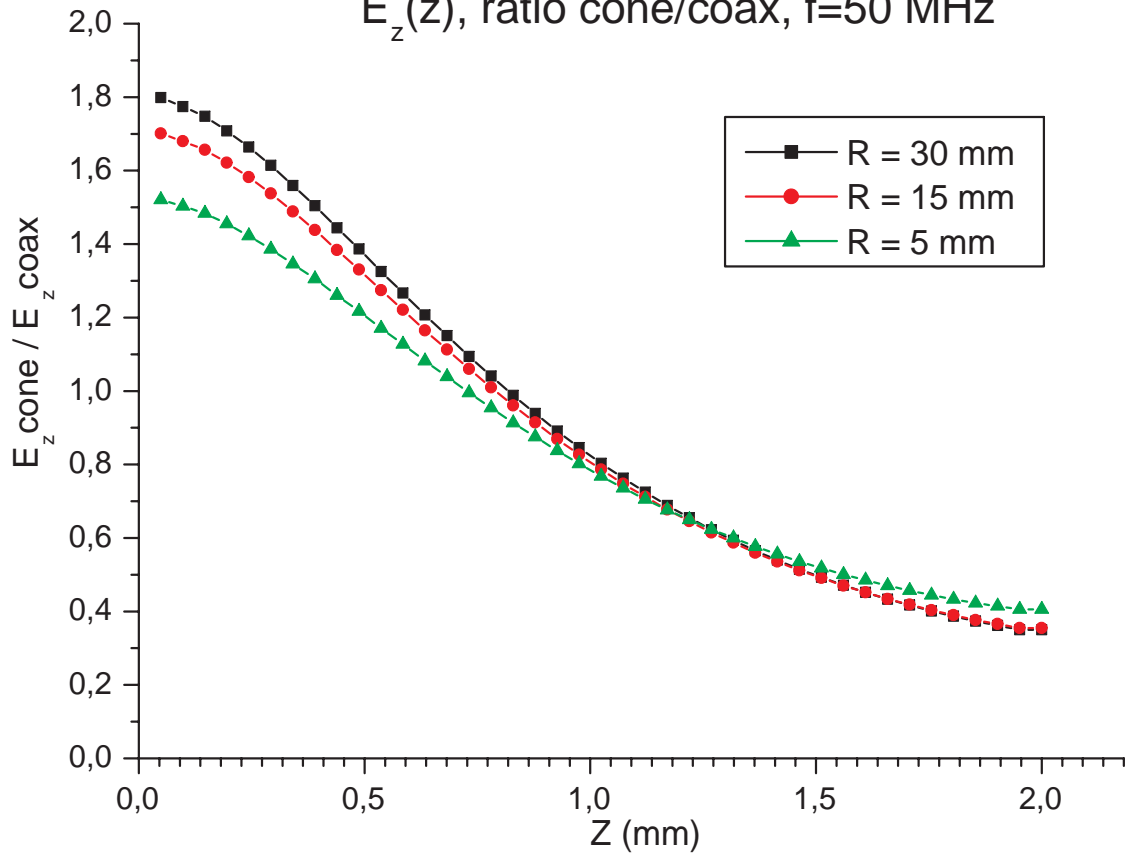

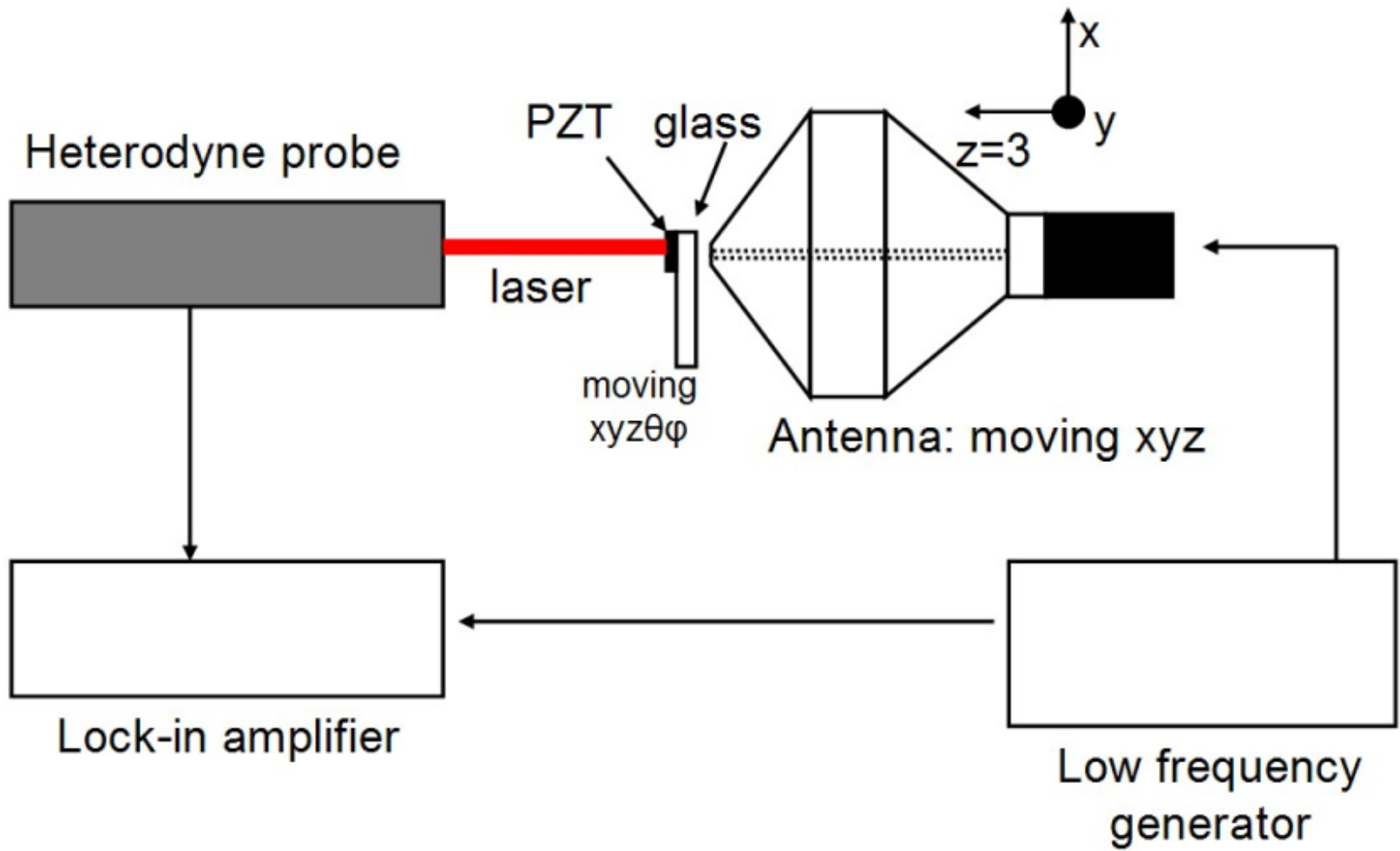

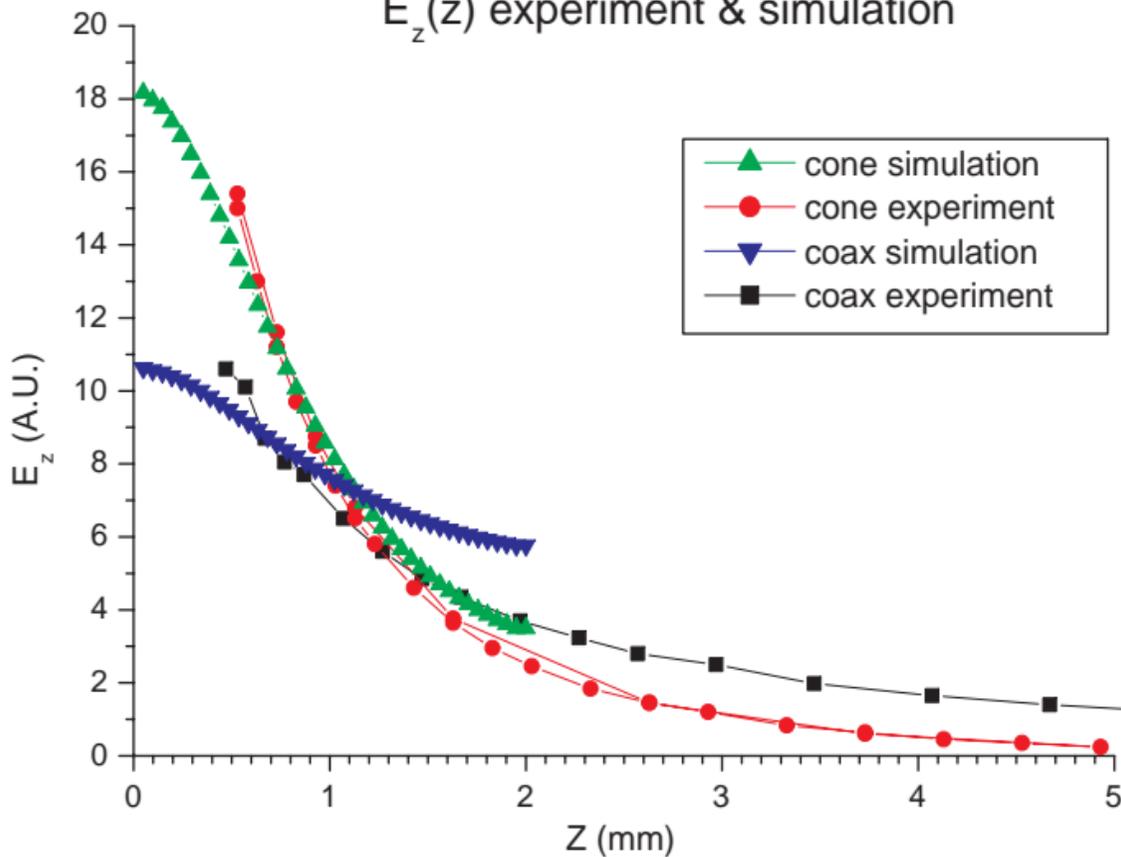